\documentclass[11pt,twoside]{article}

\usepackage{asp2004}
\usepackage{epsf}
\usepackage{lscape}
\usepackage{graphicx}
\usepackage{bm}

\begin{document}
\title{Multigrid Methods for Polarized Radiative Transfer}
\author{J. \v{S}t\v{e}p\'an}
\affil{Astronomical Institute, Academy of Sciences of the~Czech~Republic,
       CS-25165 Ond\v{r}ejov, Czech Republic}
\affil{LERMA, Observatoire de Paris-Meudon, 5 Place Jules Janssen, 
F-92195 Meudon, France}

\begin{abstract}
A new iterative method for non-LTE multilevel polarized radiative transfer
in hydrogen lines is presented. Iterative methods (such as the Jacobi method)
tend to damp out high-frequency components of the error fast, but converges
poorly due to slow reduction of low-frequency components. The idea is to use
a set of differently coarsed grids to reduce both the short- and long-period
errors. This leads to the so-called multigrid (MG) methods. For the grid of~$N$
spatial points, the number of iterations required to solve a non-LTE transfer
problem is of the order of~$O(N)$. This fact could be of great importance
for problems with fine structure and for multi-dimensional models.
The efficiency of the so-called standard MG iteration in comparison
to Jacobi iteration is shown. The formalism of density matrix is
applied to the demonstrative example of~1D, semi-infinite, non-magnetic,
3-principal level hydrogen atmospheric model. The effect of depolarizing
collisions with thermal electrons is taken into account as well as
general treatment of overlapping profiles.
\end{abstract}


\section{Formulation of the Problem}

In this paper we briefly discuss the usage of multigrid (MG) iteration 
schemes to solve the non-LTE problem of the 2nd kind as defined
by \citet{s8 La87}.
The era of extensive development of MG methods started in 1970's by 
the work of \citet{s8 Br77}.
Several steps in using MG methods applied to radiative transfer 
were made by \citet{s8 St91}, \citet{s8 Va94}, and 
\citet{s8 FB97}. These authors showed that this
technique leads to a great improvement of the convergence rate.
This paper demonstrates how to apply these methods to a more general 
solution of polarized radiative transfer with realistic multilevel 
atomic models and complicated structure of overlapping
lines. The effects of depolarizing collisions is taken into account.

For the description of the atomic state, we adopt the density-matrix 
formalism and the representation in the basis of irreducible tensorial 
operators
\citep[e.g.,][]{s8 Fa57}. The elements of atomic density matrix have 
the usual form\footnote{We suppose that all coherences between 
different energy levels vanish due to further assumptions.} 
$\rho^k_q(\alpha j)$, where $\alpha j$ is the energy level of 
total angular momentum $j$, and ($k,q$) are the multipolar components 
of the level ($k=0,\ldots,2j$, $q=-k,\ldots,k$). In stationary regime, 
the density-matrix elements are solutions of the local statistical
equilibrium equations,
\begin{equation}
\sum_{\alpha jkq} \Pi_{\alpha'j'k'q',\,\alpha jkq}\,
	\rho^k_q(\alpha j)=0\;.
\label{eq2}
\end{equation}
The structure of the $\Pi$-matrix has been extensively discussed
by \citet{s8 SB77}, \citet{s8 Bo78}, and \citet{s8 Bo80}.
We assume that this matrix has the form $\bm{\Pi}=\mathbf{R}+\mathbf{C}$, 
where $\mathbf{R}$ is the matrix of radiative rates, 
and $\mathbf{C}$ is the matrix of collisional rates (impact approximation).

The radiative transfer equation for the set of four Stokes parameters 
$\bm{S}\equiv(I,Q,U,V)^\mathrm{T}$ has the usual form
\begin{displaymath}
\frac{\mathrm{d}\bm{S}}{\mathrm{d}s}
	=\bm{J}-(\mathbf{K} - \mathbf{K}^\mathrm{s})\,\bm{S}\;,
	\label{eq1}
\end{displaymath}
where {\boldmath$J$} is the emission vector of the 
local sources, $\mathbf{K}$ is the absorption matrix, and
$\mathbf{K}^\mathrm{s}$ is the matrix of stimulated emission.
All these quantities are dependent on radiation frequency, $\nu$, 
position vector, $\bm{x}$, and direction of propagation determined 
by the unit vector $\bm{\Omega}$.  Finally, $s$ is the 
parametrization of the radiation path along the $\bm{\Omega}$ direction.

\section{Standard Multigrid Method}

Most of the existing non-LTE solvers use the methods based on
$\Lambda$-operator splitting similar to the one of
\citet{s8 RH91,s8 RH92}.
Depending on organization, these schemes are numerically
equivalent to the Jacobi or Gauss-Seidel smoothing procedures
\citep[for details, see][]{s8 PL05}.
These smoothing procedures do reduce high frequencies of the 
solution fast, but poor convergence is achieved for low frequencies. 
(With ``high frequencies'' we mean those which are comparable to 
the spatial frequency of grid points approximating the continuous 
scale.) The principles of MG schemes are based on the 
idea of using coarse grids to reduce the low frequencies,
and fine grids to smooth their high-frequency components. 
It can be showed that such a process may lead to the
optimal CPU time demands of $O(N)$, $N$ being the number of 
points per decade of optical scale. For comparison, the 
Jacobi and Gauss-Seidel methods scales approximately as 
$O(N^2)$. We have applied the non-linear version of the 
\emph{standard multigrid} scheme based on the coarse
grid correction (CGC) technique \citep[for details, see][]{s8 Ha85}.

CGC is the process of correction of the fine-grid approximation of the
solution using the solutions on the coarse grids. Schematically, it can be
described in the following way: the defect (or residuum) of the 
fine-grid approximation is computed by several calls of the sweeping 
procedure (Jacobi, etc.); then both defect and the initial guess of the
solution are restricted to the coarse grid, and a new 
solution on the coarse grid is obtained using these data.
This coarse-grid solution is interpolated to the fine grid, and 
the density-matrix components are corrected. This process can be 
repeated recursively for every grid in order of increasing grid steps.
This recursive process leads to the so-called V, W, or 
more complicated diagrams, depending on the way in which the recursion 
is implemented \citep{s8 Ha85}.

In our code we use the parabolic short-characteristics technique
of \citet{s8 KA88}, and as a smoothing algorithm we use the Jacobi 
iteration, similar to the one of \citet{s8 MT03}, with modifications 
to include the effects of line overlapping.

\section{Detailed Model Parameters}

For demonstrating the possibilities of this technique, we chose
the model of a semi-infinite, plane-parallel, \ion{H}{i} atmosphere. 
The atmosphere is isothermal ($T=5300$\,K), with constant volume
density of neutral hydrogen, $N_\mathrm{H}=10^{12}\,\mathrm{cm^{-3}}$,
thermal electrons, $N_\mathrm{e}=10^{10}\,\mathrm{cm^{-3}}$, and
protons, $N_\mathrm{p}=N_\mathrm{e}$. No magnetic field is taken into 
account. The scattering of radiation is supposed to fulfill the 
conditions of complete frequency redistribution, and the $U$ and $V$ 
Stokes parameters vanish due to the symmetries of the problem. 

We adopted a hydrogen atomic model with 9 fine-structure energy levels
($1\,\mathrm{s}^{1/2}$ to $3\,\mathrm{d}^{5/2}$) and
all the non-vanishing multipole components $\rho^k_0(nlj)$ of density matrix.
All the coherence elements ($q\ne 0$) are identically zero due to the
model's symmetries. The coherences between different energy levels 
have been neglected due to the small natural widths of the levels 
in comparison to their separations, and due to the selection rules 
for dipole radiative transitions.

Collisional rates with thermal electrons and protons for fine structure 
transitions, $C^{k\to k'}_{nlj\to n'l'j'}$, were computed in part using 
the semiclassical theory of \citet{s8 SB96}, for transitions within
the same shell, and in part using the data of collisional cross-sections
from the Atomic and Molecular Data Information System (AMDIS;
\texttt{http://www-amdis.iaea.org/}), for transitions between 
different shells.

The number of logarithmically spaced nodes in the fine grid is
$N_5=257$ and 5 grids were used in total. The number of nodes in 
the ``$G-1$'' grid is equal to $N_{G-1}=(N_{G}-1)/2+1$.
The initial guess for the atomic density-matrix elements is given 
by the LTE populations determined in the unpolarized case.

\begin{figure}[!t]
\centering
\includegraphics[width=12.1cm]{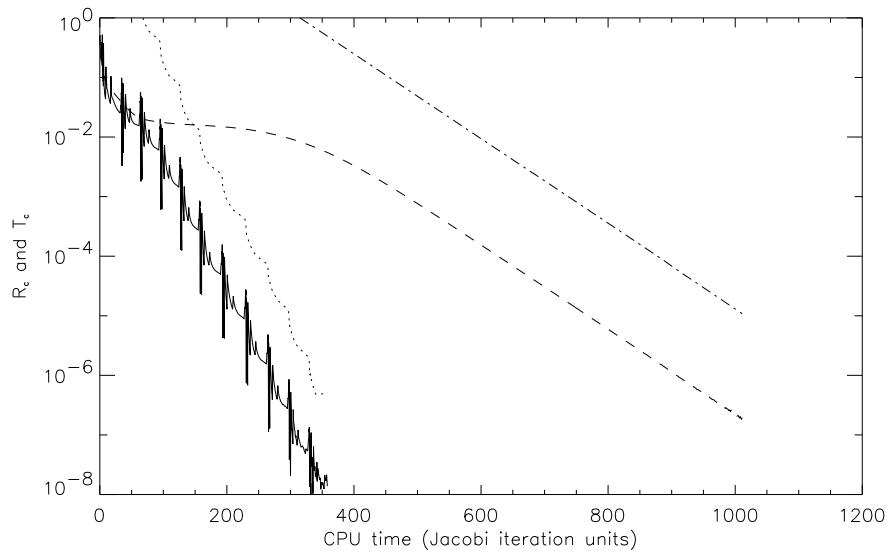}
\caption{Convergence of the MG method with Jacobi smoothing
procedure compared to the Jacobi method. The maximum relative change
$R_\mathrm{c}$ of atomic populations, $\rho^0_0$,
in the MG case (solid line) is compared to the Jacobi solution
(dashed line).
The maximum relative error $T_\mathrm{e}$ (with respect to the fully
converged solution) for MG case (dotted line) and Jacobi case 
(dot-dashed line) is showed as well. The norm used is 
$\|{\cdots}\|_\infty$ \citep*[see][]{s8 FB97}. The graph
shows the effect of 11 V-cycles with 2 pre- and 15 post-smoothing Jacobi
iterations.}
\label{fig1}
\end{figure}

\section{Convergence Properties and Conclusions}

Figure \ref{fig1} shows a comparison between the convergence rates of
the Jacobi and the MG methods. The effect of coarse-grid solutions 
is reflected
by the dramatic evolution of the maximum relative change of the
populations.  The coarser the grid the shorter the evaluation time, and 
the higher the rate of approaching to the truncation error of the grid. 
The maximum relative error dominated by the long-period components is 
strongly reduced by the recursive CGC during the V-cycles.

The time saving of the MG method in this particular model is about a 
factor 4 compared to the Jacobi method. It must be noticed that the 
efficiency of MG increases as fast as the efficiency of the smoothing 
procedure. The most important benefit from MG's is the asymptotical 
$O(N)$ behavior, which designates the method for use in solutions of 
problems that necessitate strong refinements.
Moreover the presented 1D geometry is the slowest case
as pointed out by \citet{s8 St91} and \citet*{s8 FB97}.

\acknowledgements 
I'm grateful to my Ph.D.\ supervisors, Petr Heinzel and Sylvie 
Sahal-Br\'echot, for their useful comments.

\end{document}